**KTS11-1**

# On Corrections to the Born-Oppenheimer Approximation

A Kerley Technical Services Research Report

Gerald I. Kerley, Consultant

June 2011

Kerley
Technical
Services
P.O. Box 709
Appomattox, VA 24522-0709

Website: http://kerleytechnical.com • E-mail: contact@kerleytechnical.com

# On Corrections to the Born-Oppenheimer Approximation

A Kerley Technical Services Research Report

Gerald I. Kerley, Consultant

June 2011

## ABSTRACT

This report presents a new approach for treating the coupling of electrons and nuclei in quantum mechanical calculations for molecules and condensed matter. It includes the standard "Born-Oppenheimer approximation" as a special case but treats both adiabatic and non-adiabatic corrections using perturbation theory. The adiabatic corrections include all terms that do not explicitly involve the nuclear wavefunctions, so that the nuclei move on a single electronic potential surface. The non-adiabatic corrections, which allow the nuclei to move on more than one potential surface, include coupling between the electronic and nuclear wavefunctions. The method is related to an approach first proposed by Born and Huang, but it differs in the methodology and in the definition of the electronic wavefunctions and potential surfaces. A simple example is worked out to illustrate the mechanics of the technique. The report also includes a review of previous work.

# PREFACE

Several years ago I began writing a tutorial on applications of quantum mechanics to equation-of-state theory. I had expected that a discussion of the Born-Oppenheimer approximation (BOA) would be relatively straightforward, showing it to be the leading term in a systematic perturbation expansion. (That is, after all, the general impression that one gets from textbooks.)

However, I soon learned that the theoretical framework I planned to discuss actually didn't exist. The problem of deriving the BOA and calculating corrections to it was still basically unsolved. I also discovered an enormous body of literature on the subject. As I read and studied these papers, I began to develop some new ideas of my own.

During the last six months I have taken the time to turn these ideas into a methodology that I believe to be substantially new and suitable for application to real problems in the chemistry and physics of molecules and condensed matter. This report discusses that work.

I wish to thank Brian Sutcliffe for many interesting and informative e-mails and for directing me to important sources of information. I thank George Hagedorn for the day I spent with him at Virginia Tech and for the references he provided. I thank P. K. Swaminathan for his encouragement; he too helped me out by providing some of the papers.

As always, I am especially grateful to my wonderful wife, Donna, for all her encouragement and prayers.

Finally, and most important, I thank God for helping me so much with this research project, keeping me going when I felt like quitting, and leading me to new insights and discoveries.

Gerald I. Kerley
Appomattox, VA
June 2011

P.S. The quantum mechanics tutorial still hasn't been finished, although some of the ideas were incorporated into my website presentations.

# CONTENTS

## FIGURES

# 1. INTRODUCTION

Separation of the electronic and nuclear degrees of freedom is an important problem in quantum mechanical (QM) theories of molecules and condensed matter. Most applications of QM to chemistry and physics rely on approximations to deal with the coupling between the electrons and the nuclei. Numerical calculations that include full coupling are possible only for very simple systems; even then, approximate theories are useful for interpreting the results.

## 1.1 Background

One of the first discussions of this problem appeared in the oft-cited but little-read 1927 paper by Max Born and Robert Oppenheimer (BO) [1], one year after publication of the Schrödinger Equation (SE) itself.[1] As Sutcliffe observes [5], the term "Born-Oppenheimer approximation" (BOA) is used in several different ways throughout the literature. However, it usually involves the following ideas:

> Because the electrons have much smaller masses than the nuclei, they are able to maintain an equilibrium configuration as the nuclei change positions. Therefore, one can solve the SE in two steps: First calculate the electronic structure holding the nuclear positions fixed, or "clamped." Then use the electronic energy, a parametric function of the nuclear coordinates, as the potential energy surface for the nuclear motion.

The idea that the nuclei move on a single potential surface, determined by the electronic structure, is also known as the "adiabatic approximation".[2]

The above definition is only a starting point. It does not specify what forms of the SE can be used. It does not explain *how* the disparity in the masses allows the electrons to "keep up" with the nuclei. Finally, it does not show how corrections to the approximation can be calculated and when or if they are important.

It is not surprising that this famous paper, by such distinguished authors, is sometimes taken to be the "last word," leaving nothing to be desired. Textbooks do little to challenge this perception, choosing to dwell on the merits of the idea rather than its limitations. The BOA has become so common in physics and chemistry that it is frequently used without questioning its validity, as if it were not an approximation at all. It is often assumed to be valid in virtually all QM calculations, both time-independent and time-dependent, in molecules and condensed matter.

---

1. Kutzelnigg [2] observes that Heitler and London had already anticipated the BO idea in their slightly earlier paper, a classic QM study of molecular binding [3]. BO also refer to an even earlier study by Born and Heisenberg [4], based on the "old quantum theory."
2. Note that the term "adiabatic approximation" is also used in QM problems that do not involve the BOA. See Refs. [6] and [7], for example.

The reality is much more complicated. Sutcliffe gives excellent reviews of the situation [5][8]. Hagedorn and Joye [9] discuss the problem from a mathematical point of view. Yarkony reviews non-adiabatic phenomena resulting from intersecting potential surfaces [10][11]. Kutzelnigg [2] and Essén [12] also offer interesting perspectives. The main points are summarized below.

**The BO Expansion Method**. The original BO approach had certain limitations:

- They considered only stationary states, i.e., the time-independent SE.[1]
- They considered only stable molecules (those having a configuration in which the forces on the nuclei vanish) and relatively small displacements of the nuclei from equilibrium.
- They treated the nuclear kinetic energy as a perturbation to the electronic problem and expanded the energy in a mass ratio parameter[2]—an approach quite different from the concept described above.
- The first electron-nuclear coupling term in the expansion is not particularly high-order. In fact, it appears in the same order as rotational contributions and anharmonic vibrational corrections.
- They did not provide a way to calculate the corrections or even estimate their magnitude.
- They did not allow for the effects of degeneracy, cases in which two or more potential surfaces approach or actually intersect one another.

Re-examination of the original BO mass-ratio expansion began with the work of Seiler in 1973 [15]. This approach has since been extended and placed on a more rigorous basis by mathematicians [9]. A similar expansion technique has also been applied to time-dependent problems in which the nuclear motions are represented by semi-classical wave packets [16]. However, these improvements still do not offer practical tools for computing corrections to the BOA, and they do not address phenomena that arise in cases of degeneracy and near-degeneracy.

**The Born-Huang Formalism**. In 1951, twenty-four years after the original paper,[3] Born offered the framework for a different, more general approach to treating the electron-nuclear coupling [17]. This paper was subsequently reproduced in a textbook by Born and Huang (BH) [18]. In principle, the BH method is not restricted to molecules and does not require an equilibrium nuclear configuration. It also provides for non-adiabatic effects, by writing the total wave function as a linear combination of BOA electronic wave functions. This formulation allows the nu-

1. Hagedorn and Joye [9] note that the time-dependent problem was first considered by London, in a 1928 paper that has received little attention [13]. The Landau-Zener formula for time-dependent non-adiabatic transitions appeared in 1932 [14].
2. The parameter was $\kappa = (m_e/\overline{M})^{1/4}$, $m_e$ and $\overline{M}$ being the electron and nuclear masses.
3. This delay may have been due, in part, to the momentous events that took place throughout the world during the intervening years.

clei to jump or tunnel between different electronic potential surfaces, even in a single stationary state.

Yarkony notes that non-adiabatic phenomena are now known to be much more common in molecules than previously thought [10][11]. These effects are also found in condensed matter. Born and Huang predicted that electron-nuclear coupling would be important in metals, where the electronic states form a quasi-continuum. Indeed, Ziman has described electron-phonon coupling as a non-adiabatic phenomenon [19]. Dougherty has noted that superconductivity can also be regarded as a breakdown of the BOA [20].

In contrast to the BO expansion method, the BH method has not received much attention from the mathematical community. Hagedorn, using other techniques, has made detailed studies of level crossing in time-dependent problems and has classified different types of crossings and their effects [21]. Further developments and extensions of his methods could be useful for chemical physics problems.

The BH approach has been used by a number of chemists and physicists, especially for studying non-adiabatic phenomena that arise from close approaches and intersections between potential surfaces [8][10][11]. Most of this work has studied cases of only two interacting surfaces. A formalism for solving the BH equations and applying them to a general class of problems still does not exist.

**Diabatic Transformations**. Application of the BH method to specific cases reveals that the clamped nuclei (CN) definition of the electronic potential surfaces does not always give satisfactory results. Derivatives of the electronic wavefunctions are discontinuous when potential surfaces have conic intersections, and their matrix elements are singular. This situation also leads to the geometric phase effect, non-uniqueness in the phase of the wavefunction in the vicinity of the singularities [10][11][22].

In such cases, linear combinations of the CN wavefunctions are generally used to construct so-called diabatic (or quasi-diabatic) functions that remove the singularities and some of the coupling terms. An overview of this complicated topic is beyond the scope of the present report. References [23] and [24] discuss the subject and give an extensive list of the papers relevant to it.

**Schrödinger's Cat**. Sutcliffe and Woolley [25] note that the BO and BH approximations rely on a semi-classical treatment of the nuclei. The nuclei are treated as distinguishable; permutation symmetry is rarely, if ever, fully taken into account. Hence these methods can never generate exact results and must be considered only "asymptotic" solutions at best.

It might be thought that this difficulty could be overcome by resorting to numerical calculations that treat electrons and nuclei on the same footing, without making the CN approximation [26][27]. But Sutcliffe and Woolley argue that such calculations introduce problems of their own—that they are inconsistent with conventional molecular structure concepts. In particular, they cannot distinguish between structural isomers having the same chemical formula but different chemical properties. Even the calculation of a simple dipole moment becomes problematic in such cases [28].

This issue brings to mind the dilemma of Schrödinger's cat—the unfortunate feline that is neither alive nor dead until someone has the heart to check on it [29]. The difficulty lies in the indeterminism that has always plagued quantum mechanics: QM predicts only probabilities, but we often apply it to situations where measurement selects a *specific* outcome from all *possible* outcomes. But Essén notes that this difficulty can be mitigated, in part, if the exact eigenstate is viewed as a perturbation to an approximate model having a simple interpretation [12].

**Quantum Statistical Mechanics**. The problem of electron-nuclear coupling acquires an additional dimension in equation-of-state (EOS) theories, which are based on applications of quantum *statistical* mechanics (QSM). In this case, one must also perform a *thermal* average over all energy levels of the system, including both electronic and nuclear degrees of freedom, a tremendously daunting problem even if the BOA could be used to calculate the individual states.

An approximation for simplifying the QSM problem was first presented by Zwanzig [30]. One first computes the Helmholtz free energy by making a thermal average over all electronic states, holding the nuclei fixed. This "electronic free energy" then becomes the potential surface for computing and averaging over the nuclear energy levels. This approach is currently very popular in numerical simulations, where the density functional method is used to calculate the electronic free energy [31] and the molecular dynamics method is used to calculate the nuclear motions from classical mechanics.

It is important to note that Zwanzig's approximation does *not* follow from the BOA, despite the similarity in the way it is applied. It is derived by taking the classical (high-temperature) limit of the partition function to the nuclear degrees of freedom while retaining a full quantum treatment of the electrons. Zwanzig also derived an expression for the first correction term in a high-temperature expansion. As far as I know, this correction term has never been included in any numerical calculation and is never even mentioned. There is no doubt that corrections are necessary for accurate results in some circumstances, especially at low temperatures. However, no systematic method for analyzing and calculating these corrections exists at present.

## 1.2 Overview of Report

The present study began several years ago, when I started assembling notes on quantum mechanics and its applications to EOS theory. I originally expected to show that the BOA was satisfactory for most, if not all, EOS problems. I also wanted to develop tools for deriving corrections to Zwanzig's approximation, for use in numerical EOS calculations. It soon became obvious that these issues were much more complicated than I had anticipated, that many aspects of this problem were still unsolved.

While reading the many papers on this subject and discussing it with others, I began to develop some new ideas of my own. I have spent the last six months refining these ideas and turning them into a new approach, not only for deriving the BOA, but also for extending it to include both adiabatic and non-adiabatic corrections. These ideas have now reached a point where I willing to share them with the scientific community. With further developments, they could actually be applied to problems in chemical physics.

The formalism described in this report is related to the Born-Huang approach, in that it gives the same QM description of a coupled system of electrons and nuclei. However, it differs in two important ways. It provides a systematic method for actually solving the necessary equations, something previously lacking. It also employs a new definition of the electronic wavefunctions and potential surfaces that is a better starting point for calculating the correction terms.

The remainder of this report is organized as follows:

- Sections 2 and 3 discuss definitions, notation, and theorems used herein.
- Section 4 outlines the perturbation formalism central to the method.
- Section 5, a brief digression, applies the method to a simple example.
- Section 6 discusses the new "xiabatic" electronic wavefunctions and potential surfaces.
- Section 7 gives a summary and conclusions.
- Appendices A, B, and C present additional formulas and details about equations used in the main body of the report.

I have tried to make this report readable to those who have a background in QM but are not familiar with the literature in this field. I have used simple notation and included several derivations to make the report self-contained. However it does have many equations, most of which are intermediate results used in the derivations. I have highlighted some of the most important ones by enclosing them in boxes.

# 2. BASIC CONCEPTS

## 2.1 Definition of the System

This report is concerned with solutions of the time-*independent* Schrödinger equation (SE), i.e., the stationary states

$$H(\vec{x}, \vec{X})\ \Psi_n(\vec{x}, \vec{X}) = E_n\ \Psi_n(\vec{x}, \vec{X}) . \tag{1}$$

Here $H$ is the Hamiltonian for a system of electrons and nuclei, with coordinates $\vec{x}$ and $\vec{X}$, respectively, and $\Psi_n$ and $E_n$ are the eigenfunctions and eigenvalues for all possible stationary states of the system.

I will consider only the non-relativistic case, with no spin-orbit or other correction terms. In that case, the Hamiltonian consists of kinetic energy operators $K$ and potential energy terms $U$.

$$H = K_e + K_n + U_{ee} + U_{en} + U_{nn} . \tag{2}$$

In what follows, I will designate the electronic coordinates by lower-case letters, with subscripts $i, j, \ldots$, the nuclear coordinates by upper-case letters, with Greek subscripts $\alpha, \beta, \ldots$, the nuclear charges by $Z_\alpha$, the nuclear masses by $M_\alpha$, and the electron mass by $m_e$.

The kinetic energy operators are

$$K_e = -\Sigma_i(\hbar^2/2m_e)\nabla_i^2, \quad K_n = -\Sigma_\alpha(\hbar^2/2M_\alpha)\nabla_\alpha^2 . \tag{3}$$

The electron-electron and nuclear-nuclear potential energy terms are given by the Coulomb formulas for point charges,

$$U_{ee} = \Sigma_{i>j}\, e^2/|\boldsymbol{x}_i - \boldsymbol{x}_j|, \quad U_{nn} = \Sigma_{\alpha>\beta}\, Z_\alpha Z_\beta\, e^2/|\boldsymbol{X}_\alpha - \boldsymbol{X}_\beta| . \tag{4}$$

The electron-nuclear potential term $U_{en}$ is normally defined by a similar formula. However, the assumption of point charges leads to certain problems. In particular, the electronic wavefunctions have cusps at the nuclear positions, $\boldsymbol{x}_i = \boldsymbol{X}_\alpha$, where second derivatives with respect to the nuclear positions are undefined.

These problems can be eliminated by accounting for the finite sizes of the nuclei. Appendix A derives a simple correction that is adequate for present purposes.

Treating the nuclei as spheres with radii $r_{0\alpha}$, and having a uniform charge density, one obtains

$$U_{en} = -\Sigma_i \Sigma_\alpha Z_\alpha e^2 f(r_{i\alpha}/r_{0\alpha})/r_{i\alpha}, \text{ where } r_{i\alpha} = |\boldsymbol{x}_i - \boldsymbol{X}_\alpha|, \quad (5a)$$

$$f(x) = 1 \text{ for } x \geq 1, \text{ and } f(x) = (3x - x^3)/2 \text{ for } x \leq 1. \quad (5b)$$

Equation (5a) differs from the point-charge formula only when an electron is inside a nucleus, a very rare occurrence. Appendix A shows that this expression for $U_{en}$ has finite first and second derivatives and that it eliminates the cusps in the wavefunctions.[1]

## 2.2 Other Potential Functions

Equations (4) and (5a) are the potentials corresponding to a nominally exact (nonrelativistic) description of the system. However, the perturbation method described in Sec. 4 depends only on the structure of the kinetic energy operators and can be applied to any type of potential function. Therefore, it will be useful to consider other potential functions for some problems. In Sec. 5, for example, I will show how the perturbation method can be applied to the simple example of two coupled harmonic oscillators.

It may also be of interest to consider problems where the electron-electron interactions are treated in the mean-field approximation, so that the one-electron approximation for the electronic part of the wavefunction can be used. The basic perturbation strategy will also apply to the treatment of coupling between electronic and nuclear degrees of freedom in that case.

This approach may also prove to be useful in separating the internal molecular degrees of freedom (vibrations and rotations) from the center of mass motion—an important problem in EOS theories of solids and liquids.

## 2.3 Other Coordinate Systems

Many papers discussing the BOA employ special coordinate systems—center of mass variables, rotating coordinate axes, normal modes of vibration, and sometimes even more complicated variables. While these transformations are useful for actual calculations, they are not necessary for analysis of the electron-nuclear

1. Note that this expression is not intended to give a really accurate description of the electron-nuclear potential inside the nucleus, although it is certainly more realistic than the uncorrected potential. In fact, the details of the formula will have only minor effects on the results discussed in this report. The main purpose of this exercise is to show that the numerical problems associated with the point-charge formula can be eliminated.

coupling and can actually obscure some of the essential ideas. This report will employ laboratory-fixed Cartesian coordinates except where otherwise indicated.

## 2.4 Adiabatic Basis Functions

Using the BOA as a guide, approximate wavefunctions for the total system, including both electrons and nuclei, are

$$\Psi^0_{sk}(\vec{x}, \vec{X}) = \Lambda_{sk}(\vec{X})\Phi_k(\vec{x};\vec{X}), \tag{6}$$

where $\Phi_k$ and $\Lambda_{sk}$ are wavefunctions for the electrons and nuclei, respectively. The precise definition of these functions will be postponed for the present. However, the following conditions will be required.

The $\Phi_k$ are functions of the electronic coordinates $\vec{x}$ but have a parametric dependence on the nuclear positions $\vec{X}$, indicated by a semi-colon. They are assumed to be orthonormal when integrated over the coordinates of all electrons,

$$\int \Phi_l^* \Phi_k \, d\boldsymbol{x}_1 d\boldsymbol{x}_2 \ldots \equiv \int \Phi_l^* \Phi_k \, d\vec{x} = \delta_{lk}. \tag{7}$$

They are also assumed to constitute a complete set, in that any (antisymmetric) function of the electronic coordinates can be written as an expansion of the $\Phi_k$.

The $\Lambda_{sk}$ are functions only of the nuclear coordinates $\vec{X}$. A double subscript $sk$ is used to indicate that a different set of functions can be associated with each electronic state $\Phi_k$, if so desired. Each set is assumed to be orthonormal,

$$\int \Lambda_{tk}^* \Lambda_{sk} \, d\boldsymbol{X}_1 d\boldsymbol{X}_2 \ldots \equiv \int \Lambda_{tk}^* \Lambda_{sk} \, d\vec{X} = \delta_{ts}, \tag{8}$$

The $\Lambda_{sk}$ are assumed to form a complete set with respect to functions of the nuclear coordinates. In particular, any set can be expanded in terms of any other set.

Given these conditions, it is readily seen that the $\Psi^0_{sk}$ are orthonormal,

$$\int \Psi^{0*}_{tl} \Psi^0_{sk} \, d\boldsymbol{x}_1 d\boldsymbol{x}_2 \ldots d\boldsymbol{X}_1 d\boldsymbol{X}_2 \ldots \equiv \int \Psi^{0*}_{tl} \Psi^0_{sk} \, d\vec{x} d\vec{X} = \delta_{lk}\delta_{ts}. \tag{9}$$

The adiabatic basis functions $\Psi^0_{sk}$ form a complete set, with respect to the coordinates of the electrons and nuclei. However, they cannot be exact wavefunctions for the system. Operating on $\Psi^0_{sk}$ with the Hamiltonian $H$, Eq. (1), gives

$$H\Psi^0_{sk} = \Lambda_{sk}(H^0_e\Phi_k) + \Phi_k(K_n\Lambda_{sk}) \\ -\sum_\alpha(\hbar^2/2M_\alpha)[\Lambda_{sk}\nabla^2_\alpha\Phi_k + 2\nabla_\alpha\Phi_k\cdot\nabla_\alpha\Lambda_{sk}]' \tag{10}$$

where $H^0_e$ is the so-called "clamped nuclei" (CN) Hamiltonian,

$$H^0_e = K_e + U_{ee} + U_{en} + U_{nn}. \tag{11}$$

The terms involving $\nabla^2_\alpha\Phi_k$ and $\nabla_\alpha\Phi_k$ are seen to be corrections to the BOA. If these terms were set to zero, the electronic and nuclear wavefunctions could be chosen according to the standard prescriptions

$$H^0_e\Phi_k \approx W^0_k\Phi_k \text{ and } (K_n + W^0_k)\Lambda_{sk} \approx E^0_{sk}\Lambda_{sk}. \text{ (BOA using CN)} \tag{12}$$

However, the correction terms *cannot* be made to vanish for *any* wavefunction of the adiabatic type, Eq. (6), regardless of how the $\Phi_k$ and $\Lambda_{sk}$ are defined. It is a non-trivial exercise to demonstrate when they can be neglected—and why.

It is often claimed that the electrons are able to maintain an equilibrium configuration as the nuclei change positions because their small mass allows them to move more rapidly. This concept has a certain intuitive appeal, but I am not aware of any mathematical approach that actually employs it in justifying neglect of the correction terms. In fact, Hagedorn argues that the validity of the BOA is *not* due to a disparity in *time* scales but rather to a disparity in *spatial* scales [9].

Heuristic arguments sometimes cite the use of expansions in the small parameter $m_e/M_\alpha$, (typically on the order of $10^{-6}$ to $10^{-4}$). But this point is relevant only when comparing the $\nabla^2_\alpha\Phi_k$ and $\nabla_\alpha\Phi_k$ terms to the electronic kinetic energy operator, *not* when comparing them to the nuclear kinetic energy operator. Essén has shown how the BOA can be justified using arguments that do not involve the mass ratio at all [12].

Further analysis shows that the BOA *can* be justified under certain conditions that I will discuss in subsequent sections of this report:

- The BOA will be reasonably accurate for electronic potential surfaces that are well isolated from other surfaces. Close approaches and actual intersections between surfaces lead to large non-adiabatic effects.
- The BOA should give reasonable results when the nuclei undergo only small displacements from an equilibrium configuration and do not approach any intersection points.

- The BOA should also give reasonable results for nuclear displacements that do not involve a large change in the electronic potential energy, again, provided that they do not approach intersection points.

The above conditions will generally be valid for the lowest energy levels of rotation and vibration, but they will break down for higher energy levels that involve large displacements, especially those that approach intersections between two or more potential surfaces. These higher levels play an important role in EOS modeling at high temperatures. Therefore, methods are needed for determining when corrections are important and for calculating their effects.

# 3. EXPANSIONS AND THEOREMS

## 3.1 Matrix Elements

In what follows, it will be necessary to consider matrix elements of various operators in three different basis sets. Let $\Omega$ be an operator that involves the electronic coordinates, the nuclear coordinates, or both. Matrix elements in the full basis set $\Psi^0_{sk}$ will be indicated using the so-called "bra-ket" notation:

$$\langle\, tl \mid \Omega \mid sk \,\rangle \equiv \int \Psi^{0*}_{tl}\ \Omega\ \Psi^0_{sk}\ d\vec{x}d\vec{X}. \tag{13}$$

Similar notation will be used for matrix elements involving only the nuclear functions $\Lambda_{sk}$, except with a subscript $n$ outside the ket:

$$\langle\, tl \mid \Omega \mid sk \,\rangle_n \equiv \int \Lambda^*_{tl}\ \Omega\ \Lambda_{sk}\ d\vec{X}. \tag{14}$$

Matrix elements involving only the electronic functions $\Phi_k$ will be designated with a subscript $e$:

$$\langle l|\Omega|k\rangle_e \equiv \int \Phi^*_l\ \Omega\ \Phi_k\ d\vec{x}. \tag{15}$$

Special issues arise in the case of the electronic matrix elements. First of all, they are parametric functions of the nuclear coordinates $\vec{X}$ and so cannot be assumed to commute with operators involving derivatives with respect to $\vec{X}$. Second, there is some ambiguity in the above definition when the operator $\Omega$ itself involves such derivatives. Throughout this work, I will always define the $\langle l|\Omega|k\rangle_e$ in such a way that they are *not* operators, i.e., that they *do* commute with functions of $\vec{X}$ that do not involve derivatives. (This issue will become clearer when specific cases are addressed.)

## 3.2 Expansions

As noted above, the $\Psi^0_{sk}$ are a complete set of functions on the full coordinate space $\{\vec{x}, \vec{X}\}$, while the $\Phi_k$, and $\Lambda_{sk}$ are complete on the subspaces $\{\vec{x}\}$ and $\{\vec{X}\}$, respectively. Let $\psi_n(\vec{\xi})$ denote the wavefunctions and coordinates for any one of the above sets. Given an operator $\Omega$, completeness allows one to expand

$$\Omega\psi_n(\vec{\xi}) = \Sigma_m \langle n|\Omega|m\rangle \psi_m(\vec{\xi}), \tag{16}$$

where the sum is taken over all functions in the set.

The above result can be used to derive a sum rule for the product of two operators. Consider the *special case* of an operator $\Omega_2$ that commutes with the matrix elements of another operator $\Omega_1$,

$$\Omega_2\langle n|\Omega_1|m\rangle - \langle n|\Omega_1|m\rangle\Omega_2 = 0. \tag{17}$$

If this relation holds, Eq. (16) gives

$$\Omega_2\Omega_1\psi_m(\vec{\xi}) = \Sigma_p\Omega_2\langle m|\Omega_1|p\rangle\psi_p(\vec{\xi}) = \Sigma_p\langle m|\Omega_1|p\rangle\Omega_2\psi_p(\vec{\xi}); \tag{18}$$

Multiplying by $\psi_n^*(\vec{\xi})$ and integrating, one finds

$$\langle n|\Omega_2\Omega_1|m\rangle = \Sigma_p\langle n|\Omega_2|p\rangle\langle p|\Omega_1|m\rangle, \text{ [}\textit{only}\text{ if Eq. (17) holds].} \tag{19}$$

Note that this result will *not* hold in certain cases. Matrix elements of the electronic wavefunctions $\Phi_k$ are parametric functions of the nuclear coordinates and so will not commute with an operator that involves derivatives with respect to those coordinates. Special handling is required in such cases.

An operator $\Omega$ is said to be *Hermitian* if $\langle n|\Omega|m\rangle = \langle m|\Omega|n\rangle^*$. QM Hamiltonian operators are normally (but not always) Hermitian [35].

If the $\psi_n$ are eigenfunctions of some (Hermitian) Hamiltonian $H$, then

$$H\psi_n(\vec{\xi}) = \varepsilon_n\psi_n(\vec{\xi}) \text{ and } \langle n|H|m\rangle \equiv H_{nm} = \varepsilon_n\delta_{nm}. \tag{20}$$

The eigenfunctions of $H$ can be expanded in terms of an approximate (but complete) basis set $\psi_n^0$. The expansion coefficients and eigenvalues can then be computed, at least in principle, by matrix diagonalization, perturbation theory, or a combination of the two methods. (Reference [35] discusses perturbation theory techniques, including some methods that are not found in QM textbooks.)

### 3.3 The Commutator Theorem

The above equations can be used to derive a useful theorem. If the basis functions $\psi_n$ are eigenfunctions of a Hamiltonian $H$, the sum rule, Eq. (19), gives

$$\langle n|H\Omega|m\rangle = \Sigma_p\langle n|H|p\rangle\langle p|\Omega|m\rangle = \varepsilon_n\langle n|\Omega|m\rangle, \tag{21}$$

*provided* $H$ commutes with the matrix elements $\langle n|\Omega|m\rangle$.

Reversing the operators $H$ and $\Omega$, one obtains

$$\langle n|\Omega H|m\rangle = \langle n|\Omega \varepsilon_m|m\rangle = \varepsilon_m\langle n|\Omega|m\rangle + \langle n|\Omega\varepsilon_m - \varepsilon_m\Omega|m\rangle. \quad (22)$$

Note that the above result allows for cases where $\Omega$ does not commute with $\varepsilon_m$. This situation occurs when one is calculating matrix elements of the electronic wavefunctions $\Phi_k$ and the operator $\Omega$ involves derivatives with respect to the nuclear coordinates.

Combining Eqs. (21) and (22), one obtains

$$\boxed{\langle n|[H, \Omega]|m\rangle = (\varepsilon_n - \varepsilon_m)\langle n|\Omega|m\rangle - \langle n|\Omega\varepsilon_m - \varepsilon_m\Omega|m\rangle}, \quad (23)$$

where $[H, \Omega] \equiv H\Omega - \Omega H$. I will call this equation the *commutator theorem*.[1]

Note that Eq. (23) is valid for an electronic Hamiltonian that depends parametrically on the nuclear configuration, *provided* that it does not involve any derivatives with respect to the nuclear coordinates.

### 3.4 The Hellmann-Feynman Theorem

Equation (23) can be used to derive the well-known Hellmann-Feynman theorem [37]. If $H$ is a parametric function of a parameter $\lambda$, and $\Omega = \partial/\partial\lambda$, one obtains

$$\langle n|\partial H/\partial\lambda|m\rangle = (\varepsilon_m - \varepsilon_n)\langle n|(\partial/\partial\lambda)|m\rangle + (\partial\varepsilon_m/\partial\lambda)\delta_{mn}. \quad (24)$$

Note that Eq. (24) includes both diagonal and off-diagonal matrix elements of the operators $\Omega$ and $\partial H/\partial\lambda$. The diagonal case reduces to the formula usually presented in textbooks.

Inspection of the off-diagonal terms, together with Eq. (16), reveals that problems arise when there are degenerate states. The derivatives $\partial\psi_m/\partial\lambda$ are undefined if $\langle n|\partial H/\partial\lambda|m\rangle \neq 0$ for $\varepsilon_m = \varepsilon_n$ and $m \neq n$. That is why the Hellmann-Feynman theorem is sometimes found to give erroneous results when used to calculate the forces on nuclei for degenerate electronic states. Correct results require diagonalization of the force matrix [38].

Degenerate states, i.e., overlapping potential surfaces, also lead to problems in the matrix elements of $\nabla_\alpha$ and $\nabla_\alpha^2$ that arise in the BOA corrections.

1. Despite the simplicity and usefulness of this result, which I first used in my thesis research [36], I have never found a presentation of it in any QM textbook or paper.

# 4. PERTURBATION METHOD

## 4.1 Overview

As noted in Sec. 2, the adiabatic basis functions $\Psi^0_{sk}$, Eq. (6), form a complete set with respect to the coordinates of the electrons and nuclei. Therefore, the exact eigenfunctions of the Hamiltonian can be expressed by the expansion,

$$\Psi_n(\vec{x}, \vec{X}) = \Sigma_{sk} C(sk;n) \Psi^0_{sk}(\vec{x}, \vec{X}) = \Sigma_{sk} C(sk;n) \Lambda_{sk}(X) \Phi_k(\vec{x}; \vec{X}), \quad (25)$$

where the expansion coefficients $C(sk;n)$ can, in principle, be determined by perturbation theory and/or matrix diagonalization.

Application of this approach involves three tasks that will be discussed in this and following sections:

- definition of the electronic and nuclear basis functions, $\Phi_k$ and $\Lambda_{sk}$,
- calculation of the matrix elements, and
- a methodology for solving the perturbation equations.

## 4.2 Relation to Born-Huang Approach

Before addressing the above tasks, note that Eq. (25) can be rewritten by summing the expansion coefficients over all the nuclear basis functions, as follows:

$$\Psi_n(\vec{x}, \vec{X}) = \Sigma_k \chi_{nk}(\vec{X}) \Phi_l(\vec{x}; \vec{X}), \text{ where} \quad (26a)$$

$$\chi_{nk}(\vec{X}) = \Sigma_s C(sk;n) \Lambda_{sk}(\vec{X}). \quad (26b)$$

Equation (26a) has exactly the same form as that proposed by Born and Huang (BH) and so provides the same physical description of the system. It is explicitly non-adiabatic, allowing the nuclei to move on more than one potential surface, the coefficients $\chi_{nk}$ determining the contributions of the electronic states as functions of the nuclear coordinates.

The method presented here is formally equivalent to BH, but it differs in two important ways. First, it offers a systematic methodology for determining the expansion coefficients, something that has been lacking until now. Second, the electronic and nuclear basis functions will be chosen by different rules that provide a better starting point for the perturbation calculations.

### 4.3 Electronic Matrix Elements

Let us now consider the matrix elements of the full Hamiltonian $H$ in the adiabatic basis set. Multiplying Eq. (10) by $\Phi_l^*$, and integrating over the electronic coordinates, one obtains

$$\int \Phi_l^* H \Psi_{sk}^0 d\vec{x} = [K_n \delta_{kl} + \langle l | H_e^0 | k \rangle_e + u_{lk} + 2\omega_{lk}] \Lambda_{sk}(\vec{X}), \tag{27}$$

where

$$u_{lk} = -\Sigma_\alpha (\hbar^2/2M_\alpha) \langle l | \nabla_\alpha^2 | k \rangle_e, \tag{28}$$

$$\boxed{\omega_{lk} = -\Sigma_\alpha (\hbar^2/2M_\alpha) \langle l | \nabla_\alpha | k \rangle_e \cdot \nabla_\alpha}\,. \tag{29}$$

The BH formulation chooses the electronic functions $\Phi_k$ to be eigenfunctions of the CN Hamiltonian $H_e^0$, Eq. (11), making $\langle l | H_e^0 | k \rangle_e = W_k^0 \delta_{kl}$ a diagonal matrix. A slight improvement is sometimes made by adding the diagonal correction term $u_{kk}$ to $W_k^0$ when computing the nuclear functions. This correction is generally small, but it can have a significant effect in some situations [39].

The off-diagonal terms $u_{lk}$ and $\omega_{lk}$ are usually regarded as "non-adiabatic" corrections. However, one can improve upon this approach by including some of the *off*-diagonal $u_{lk}$ terms in the electronic wavefunctions and potential surfaces.

First derive some relations involving the electronic matrix elements. Since the overlap conditions $\langle l \mid k \rangle_e = \delta_{kl}$ are valid for all nuclear configurations, one has

$$\langle l | \nabla_\alpha | k \rangle_e = \nabla_\alpha \langle l \mid k \rangle_e - \langle k | \nabla_\alpha | l \rangle_e^* = -\langle k | \nabla_\alpha | l \rangle_e^*, \tag{30}$$

$$\omega_{lk} = -\omega_{kl}^*. \tag{31}$$

Now note that

$$\langle l | \nabla_\alpha^2 | k \rangle_e = \nabla_\alpha \cdot \langle l | \nabla_\alpha | k \rangle_e - \int (\nabla_\alpha \Phi_l^*) \cdot (\nabla_\alpha \Phi_k) d\vec{x}, \tag{32}$$

$$\langle k | \nabla_\alpha^2 | l \rangle_e^* = \nabla_\alpha \cdot \langle k | \nabla_\alpha | l \rangle_e^* - \int (\nabla_\alpha \Phi_l^*) \cdot (\nabla_\alpha \Phi_k) d\vec{x}, \tag{33}$$

$$\int(\nabla_\alpha\Phi_l^*)\cdot(\nabla_\alpha\Phi_k)d\vec{x} = -\tfrac{1}{2}(\langle l|\nabla_\alpha^2|k\rangle_e + \langle k|\nabla_\alpha^2|l\rangle_e^*). \qquad (34)$$

$u_{lk}$ can now written in terms of a Hermitian component and a remainder,

$$u_{lk} = \nu_{lk} + \Delta u_{lk}, \text{ where} \qquad (35)$$

$$\boxed{\begin{aligned}\nu_{lk} &= \textstyle\sum_\alpha(\hbar^2/2M_\alpha)\int(\nabla_\alpha\Phi_l^*)\cdot(\nabla_\alpha\Phi_k)d\vec{x} \\ &= -\textstyle\sum_\alpha(\hbar^2/2M_\alpha)\,\tfrac{1}{2}[\langle l|\nabla_\alpha^2|k\rangle_e + \langle k|\nabla_\alpha^2|l\rangle_e^*]\end{aligned}}, \qquad (36)$$

$$\Delta u_{lk} = -\textstyle\sum_\alpha(\hbar^2/2M_\alpha)\nabla_\alpha\cdot\langle l|\nabla_\alpha|k\rangle_e. \qquad (37)$$

I now propose to define the electronic functions by requiring

$$\boxed{\langle l|H_e^0|k\rangle_e + \nu_{lk} = W_k\delta_{kl}}. \qquad (38)$$

Note that the Hermitian matrix $\nu_{lk}$ is used here, instead of $u_{lk}$, so that the electronic wavefunctions will be orthogonal. Using this result, Eq. (27) becomes

$$\int\Phi_l^* H\,\Psi_{sk}^0 d\vec{x} = [(K_n + W_k)\delta_{lk} + \Delta u_{lk} + 2\omega_{lk}]\,\Lambda_{sk}(\vec{X}). \qquad (39)$$

The solution to Eq. (38), if it can be found, is a significant improvement to the CN potential surface. Because it does not involve the nuclear wavefunctions, it includes all BOA corrections that retain the "adiabatic" picture—the nuclei moving on a single potential surface. I will refer to the electronic wavefunctions as an *extended* adiabatic basis, or "xiabatic" basis set.[1] This xiabatic potential surface should be sufficiently accurate for calculating the nuclear wavefunctions in configurations where the non-adiabatic terms $\Delta u_{lk} + 2\omega_{lk}$, are small.

This new xiabatic basis is *not* equivalent to the well-known diabatic (or quasi-diabatic) basis that is used to eliminate (or minimize) the $\omega_{lk}$ coupling terms [10][11][23][24]. However, it may have some of the same advantages. This matter is discussed further in Sec. 6.

1. This xiabatic basis set is "extended" only in the sense that it goes beyond the traditional clamped nuclei basis set. Different wavefunctions are still calculated at each point on the potential surface, and the basis set is not overdetermined.

I have not yet found a way to solve Eq. (38) exactly, or even to prove that an exact solution exists. However, Sec. 6 outlines an approximate solution that should be adequate for many purposes and provide a starting point for improvements.

### 4.4 Nuclear Matrix Elements

Next multiply Eq. (39) by $\Lambda_{tl}^{*}$, and integrate over the nuclear coordinates, to obtain the Hamiltonian matrix elements in the adiabatic basis set. The result is

$$\langle\, tl \mid H \mid sk \,\rangle = \langle\, tk \,|K_n + W_k|\, sk \,\rangle_n \delta_{lk} + \langle\, tl \,|\Delta u_{lk} + 2\omega_{lk}|\, sk \,\rangle_n . \tag{40}$$

The $\Delta u_{lk}$ can be expressed in terms of the $\omega_{lk}$, as follows. Note that

$$\begin{aligned} \langle\, tl \mid \nabla_\alpha \cdot \langle l \,|\nabla_\alpha|\, k \rangle_e \mid sk \,\rangle_n = & \int \nabla_\alpha \cdot [\Lambda_{tl}^{*} \langle l \,|\nabla_\alpha|\, k \rangle_e \Lambda_{sk}]\, d\tilde{X} \\ & - \langle\, tl \mid \langle l \,|\nabla_\alpha|\, k \rangle_e \cdot \nabla_\alpha \mid sk \,\rangle_n \quad , \\ & - \langle\, sk \mid \langle l \,|\nabla_\alpha|\, k \rangle_e^{*} \cdot \nabla_\alpha \mid tl \,\rangle_n^{*} \end{aligned} \tag{41}$$

where the first term on the right-hand side is seen to vanish. Using this result, together with Eqs. (29)-(31) and (37), one finds

$$\langle\, tl \,|\Delta u_{lk}|\, sk \,\rangle_n = -\langle\, tl \,|\omega_{lk}|\, sk \,\rangle_n + \langle\, sk \,|\omega_{kl} \mid tl \,\rangle_n^{*} . \tag{42}$$

Hence the Hamiltonian matrix elements, in the adiabatic basis set, are

$$\begin{aligned} \langle\, tl \mid H \mid sk \,\rangle = & \langle\, tk \,|K_n + W_k|\, sk \,\rangle_n \delta_{lk} \\ & + \langle\, tl \,|\omega_{lk}|\, sk \,\rangle_n - \langle\, sk \,|\omega_{kl}|\, tl \,\rangle_n^{*} . \end{aligned} \tag{43}$$

Switching indices, and applying Eq. (31), it is easily verified that $\langle\, tl \mid H \mid sk \,\rangle$ is a Hermitian matrix, as required.

I now propose to choose the nuclear wavefunctions from the eigenvalue equation

$$\boxed{(K_n + W_k)\Lambda_{sk} = E_{sk}^{0}\Lambda_{sk}} . \tag{44}$$

Finally, the matrix elements of the Hamiltonian, in the adiabatic basis set, are

$$\boxed{\langle\, tk \mid H \mid sk \,\rangle = (E_{sk}^{0} + 2\omega_{kk})\delta_{st}} , \tag{45}$$

$$\boxed{\langle\, tl \,|\, H \,|\, sk \,\rangle \;=\; \langle\, tl \,|\omega_{lk}|\, sk \,\rangle_n - \langle\, sk \,|\omega_{kl}|\, tl \,\rangle_n^*,\;\; (k \neq l)}\,. \tag{46}$$

(Note that the $\omega_{kk}$ are imaginary and vanish if the $\Phi_k$ are real functions. These terms are included here for generality, in case it is not possible or desirable to use real functions.)

## 4.5 Summary of the Method

The perturbation method developed thus far uses the following procedure for solving the Schrödinger equation for a system of electrons and nuclei:

- First find the electronic wavefunctions that satisfy Eq. (38), i.e., by diagonalizing $\langle l|H_e^0|k\rangle + v_{lk}$, which I will call the "xiabatic matrix."
- Next solve Eq. (44) for the nuclear wavefunctions and energies, using the diagonal electronic matrix elements $W_k$ as potential functions.
- Finally, compute the full Hamiltonian matrix from Eqs. (45) and (46), and use perturbation theory to compute the non-adiabatic corrections.

This approach allows one to determine whether the BOA is or is not accurate and to calculate the necessary corrections when it is not.

Note that the method identifies *two* kinds of corrections to the BOA. The xiabatic terms $v_{lk}$, Eq. (36), affect the electronic wavefunctions and potential surfaces but do not couple the electronic and nuclear degrees of freedom. The non-adiabatic terms $\omega_{lk}$, Eq. (29), generate off-diagonal coupling terms that allow the nuclei to move between different electronic surfaces.

Both types of corrections involve derivatives of the electronic wavefunctions with respect to the nuclear coordinates. Section 6 and Appendix A use the Hellmann-Feynman theorem to relate the matrix elements of these derivatives are to those involving derivatives of the electron-nuclear potential energy. These corrections will be small as long as the nuclear motions are confined to regions where the potential surface is well isolated from other potential surfaces.

Those conditions will often be satisfied for small, quasi-harmonic displacements of the nuclei from an equilibrium condition—the case originally considered by BO. However, they can also apply to nuclear motions that are *not* harmonic—such as internal rotations within molecules and the motions of molecules in liquids—especially when they do not involve large changes in the potential energy.

The xiabatic corrections $v_{lk}$ have roughly the same functional form as the electronic kinetic energy but are much smaller by the electron-nuclear mass ratio. Hence these terms will have only a small effect on the clamped-nuclei potential surface in regions where it is well separated from other surfaces. But corrections to the CN surface will be important for configuations where two or more surfaces approach or actually overlap.

The non-adiabatic terms will also be most important for configurations where two or more surfaces approach or overlap. However, they give rise to a different effect—allowing the nuclei to tunnel and move between the potential surfaces. In this case, the corrections also involve the nuclear wavefunctions and can be calculated by perturbation theory. The non-adiabatic effects will have much less effect for nuclear motions restricted to configurations where the surfaces are well separated, although they will never vanish completely.

I proposed above that the nuclear wavefunctions $\Lambda_{sk}$ be eigenvalues of the effective nuclear Hamiltonians $K_n + W_k$, Eq. (29). This procedure, which leads to a different set of nuclear functions for each potential surface, should be most reasonable for configurations in which the nuclear motions are restricted to regions where the potential surface $W_k$ is well-separated from other surfaces.

However, that approach may not be tractable for highly-excited nuclear states, especially those involving nuclear configuations where the coupling operators $\omega_{lk}$ are large. In such cases, it might be useful to modify the potential surfaces so that similar nuclear wavefunctions are obtained on adjacent surfaces. This "restructuring" of the potential surfaces might facilitate calculation of the wavefunctions themselves as well as the coupling matrix elements. Application of the method to specific problems should suggest ways to proceed.

The next section will demonstrate the use of the methodology developed thus far on a simple example. Further discussion of the electronic problem will be given in Sec. 6.

# 5. A SIMPLE EXAMPLE

The equations developed in the previous section depend only on the form of the kinetic energy operators and make no assumptions about the potential energy. This section illustrates the mechanics of the method by applying it to the simple example of two coupled harmonic oscillators, a problem posed by Fernández [40][41]. Discussion of the electron-nuclear problem resumes in Sec. 6.

## 5.1 Definition of Problem

The Hamiltonian, in terms of dimensionless variables, is

$$H = -\tfrac{1}{2}\partial^2/\partial x^2 - \tfrac{1}{2}\beta\partial^2/\partial X^2 + \tfrac{1}{2}(x^2 + X^2 + 2\lambda xX), \tag{47}$$

where $\beta$ is the mass ratio and $\lambda$ is a coupling constant. With $\beta < 1$, $x$ and $X$ are the coordinates of the lighter and heavier oscillators, respectively. The exact energy levels of this system depend upon two quantum numbers [40][41]:

$$E_{ks} = (k + \tfrac{1}{2})\sqrt{\varepsilon_+} + (s + \tfrac{1}{2})\sqrt{\varepsilon_-}, \text{ where} \tag{48a}$$

$$\varepsilon_\pm = \tfrac{1}{2}(1 + \beta \pm \sqrt{(1-\beta)^2 + 4\lambda^2\beta}). \tag{48b}$$

Notice that the above solution exists only for $\lambda \le 1$, since $\varepsilon_- < 0$ for $\lambda > 1$.

Despite its simplicity, this system has a number of features that resemble the electron-nuclear systems of interest in this work. It has two sets of energy levels. Those for the lighter oscillator, with coordinate $x$, are more widely spaced, as electronic surfaces normally are. The levels of the heavier oscillator, with coordinate $X$, are more closely spaced, like the vibrational-rotational states of the nuclei. And, as shown below, treatment of the coupling between these two sets of levels requires both adiabatic and non-adiabatic corrections to the BOA.

The harmonic-oscillator relations used in this section are given in Appendix B.

## 5.2 Eigenstates of First Oscillator

Let $H_1^0$ be the Hamiltonian obtained by holding coordinate $X$ fixed:

$$\begin{aligned} H_1^0 &= -\tfrac{1}{2}\,\partial^2/\partial x^2 + \tfrac{1}{2}(x + \lambda X)^2 + \tfrac{1}{2}(1-\lambda^2)X^2 \\ &= h_0(x + \lambda X) + \tfrac{1}{2}(1-\lambda^2)X^2 \end{aligned}, \tag{49}$$

where $h_0$ is the harmonic-oscillator Hamiltonian, Eq. (B.1). The last term does not depend on $x$ and can be treated as additive in calculating the eigenfunctions of the first oscillator.

Solving for the eigenfunctions of $H_1^0$ would be equivalent to making the clamped nuclei (CN) approximation in the electron-nuclear problem, giving energy levels that are independent of $\beta$ and $\lambda$. Some improvement could be obtained by including the diagonal correction term $v_{kk}$, Eq. (36).

The present treatment will go even further to obtain the xiabatic solution by solving Eq. (38) with the adiabatic corrections $v_{lk}$. First require the wavefunctions for oscillator 1 to satisfy the eigenvalue problem

$$H_1\Phi_k(x;X) = W_k^0\Phi_k(x;X), \text{ where } H_1 = H_1^0 + \tfrac{1}{2}a(x+\lambda X)^2, \tag{50}$$

and $a$ is a constant to be determined. A substitution of variables gives

$$H_1 = \sqrt{1+a}\; h_0(\xi) + \tfrac{1}{2}(1-\lambda^2)X^2, \text{ where} \tag{51a}$$

$$\xi = b(x+\lambda X) \text{ and } b = (1+a)^{1/4}. \tag{51b}$$

Using the equations given in Appendix B, the eigenvalues and eigenfunctions are

$$W_k^0(X) = (k+\tfrac{1}{2})\sqrt{1+a} + \tfrac{1}{2}(1-\lambda^2)X^2 \text{ and} \tag{52a}$$

$$\Phi_k(x;X) = b^{1/2}\psi_k(\xi), \tag{52b}$$

where $\psi_k(\xi)$ is a harmonic oscillator wavefunction. Note that the factor $b^{1/2}$, in Eq. (52b), is needed to normalize the functions in terms of the variable $x$, instead of $\xi$. In what follows, the subscript $1$ will be used for matrix elements when $x$ is the integration variable, the subscript $0$ when $\xi$ is the integration variable.

In order to satisfy Eq. (38), one must have

$$\langle l\,|H_1^0|\,k\rangle_1 + \nu_{lk} = \langle l\,|H_1|\,k\rangle_1 - \tfrac{1}{2}(a/b^2)\langle l\,|\xi^2|\,k\rangle_0 + \nu_{lk} = W_k\delta_{kl}. \tag{53}$$

This condition can be satisfied by an appropriate choice of the constant $a$. For the present problem, Eq. (36) has the form

$$\nu_{lk} = \tfrac{1}{2}\beta\int_{-\infty}^{\infty}(\partial\Phi_l/\partial X)(\partial\Phi_k/\partial X)dx. \tag{54}$$

Noting that

$$\partial\Phi_k/\partial X = (\partial\xi/\partial X)(\partial\Phi_k/\partial\xi) = b^{3/2}\lambda(d\psi_k/\partial\xi), \tag{55}$$

and making use of Eq. (B.12), one obtains

$$\nu_{lk} = \beta b^2\lambda^2(\langle l|h_0|k\rangle_0 - \tfrac{1}{2}\langle l|\xi^2|k\rangle_0) = \beta b^2\lambda^2[(k+\tfrac{1}{2})\delta_{lk} - \tfrac{1}{2}\langle l|\xi^2|k\rangle_0]. \tag{56}$$

The diagonal requirement, Eq. (53), is satisfied by requiring

$$a = -\beta\lambda^2/(1+\beta\lambda^2). \tag{57}$$

The final result is

$$W_k = W_k^0 + \beta b^2\lambda^2(k+\tfrac{1}{2}) = (k+\tfrac{1}{2})\sqrt{1+\beta\lambda^2} + \tfrac{1}{2}(1-\lambda^2)X^2. \tag{58}$$

### 5.3 Eigenstates of Second Oscillator

The next step is to solve Eq. (44) for the eigenvalues and eigenfunctions of the second oscillator. The operator $\omega_{kk}$ vanishes because the $\Phi_k$ are real, so $W_k$ is the potential function for the motion. Since the $k+\frac{1}{2}$ term in Eq. (58) is independent of the $X$ coordinate, all potential surfaces $W_k$ are parallel and give the same set of eigenfunctions, $\Lambda_{sk} \rightarrow \Lambda_s$. Define the Hamiltonian for oscillator 2 as

$$H_2 = -\tfrac{1}{2}\beta\, d^2/dX^2 + \tfrac{1}{2}(1-\lambda^2)X^2 = \sqrt{\beta(1-\lambda^2)}\; h_0(z). \tag{59}$$

where $z = [(1-\lambda^2)/\beta]^{1/4}X$. The eigenvalues and eigenfunctions of $H_2$ are

$$H_2\Lambda_s(X) = (s+\tfrac{1}{2})\sqrt{\beta(1-\lambda^2)}\;\Lambda_s(X) \text{ and} \tag{60a}$$

$$\Lambda_s(X) = [(1-\lambda^2)/\beta]^{1/8}\psi_s(z). \tag{60b}$$

### 5.4 Adiabatic Result

Hence the diagonal elements of the Hamiltonian matrix, Eq. (45), are

$$\langle\, sk \,|\, H \,|\, sk \,\rangle = E_{sk}^0 = (k+\tfrac{1}{2})A + (s+\tfrac{1}{2})B, \text{ where} \tag{61a}$$

$$A = \sqrt{1+\beta\lambda^2},\; B = \sqrt{\beta(1-\lambda^2)}. \tag{61b}$$

This result is the best estimate of the energy levels that can be made without including the non-adiabatic corrections. Comparing it with the exact result, Eq. (48a), shows that the expressions corresponding to Eq. (48b) are exact to leading order in $\beta$ but omit higher-order terms.

Figures 1a and 2a show the ground-state energy, $k = s = 0$, as a function of the mass ratio $\beta$, for two values of the coupling constant, $\lambda = 0.5$ and $\lambda = 0.9$. The adiabatic result, shown in red, agrees quite well with the exact result, shown by circles, for values of $\beta$ up to ~0.1. The adiabatic curve becomes less accurate at higher values, overestimating the total energy by up to 5% at $\beta = 1$.

Figures 1b and 2b show the contributions from the individual oscillators to the ground state energy. Here again, the agreement between the adiabatic and exact results is good up to $\beta = 0.1$. At higher values of $\beta$, the errors in the two terms have opposite signs and tend to cancel one another. However, the individual contributions are also important because they determine the spacings between excitations. The adiabatic approximation underestimates the excitation energies of oscillator 1 and overestimates the excitation energies of oscillator 2.

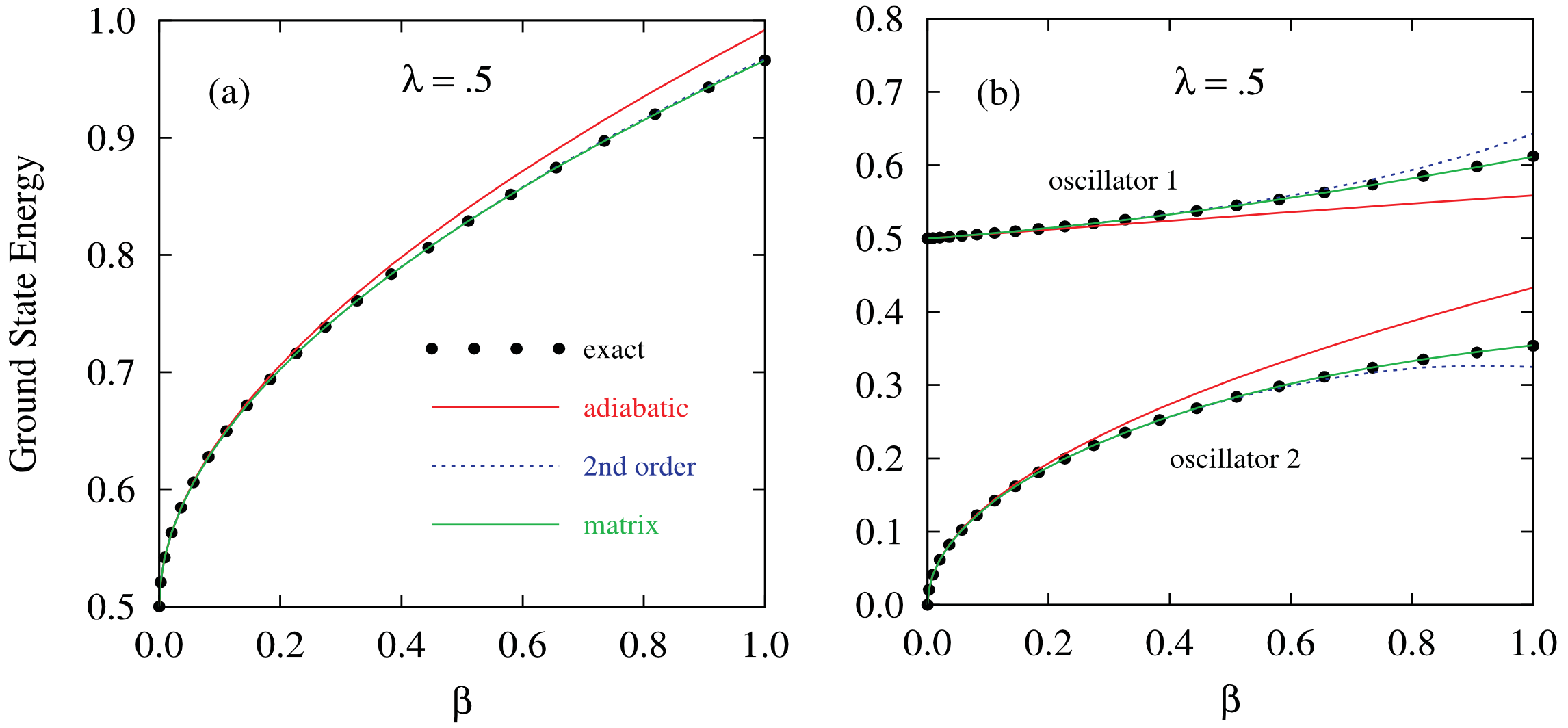

**Fig. 1. Ground state energy vs. mass-ratio β, for coupled oscillators with λ = 0.5: (a) total energy; (b) contributions from oscillators 1 and 2. Circles—exact results, Eq. (48a); red curve—adiabatic results, Eq. (61a); blue curve—includes 2nd-order non-adiabatic corrections, Eq. (66); green curve—results from full matrix.**

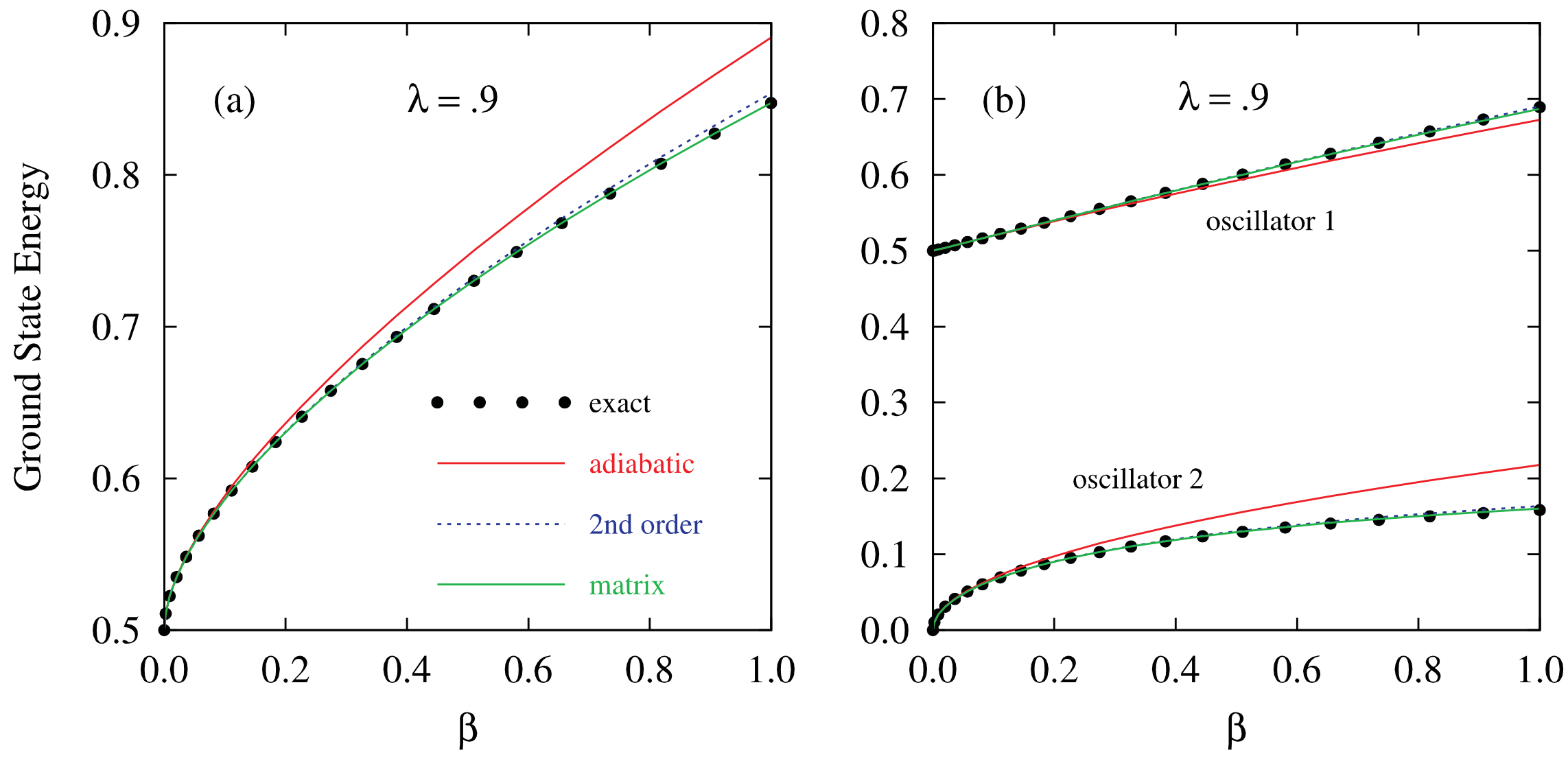

**Fig. 2. Ground state energy vs. mass-ratio β, for coupled oscillators with λ = 0.9: (a) total energy; (b) contributions from oscillators 1 and 2. Curves are same as in Fig. 1.**

## 5.5 Non-Adiabatic Corrections

Further improvements require calculation of the non-adiabatic corrections, using the off-diagonal matrix elements, Eq. (46).

First consider the operator $\omega_{lk}$, Eq. (29). For the present example,

$$\omega_{lk} = -\tfrac{1}{2}\beta\langle l|\partial/\partial X|k\rangle_1 \partial/\partial X = -\tfrac{1}{2}\beta b\lambda\langle l|\partial/\partial\xi|k\rangle_0 \partial/\partial X, \tag{62}$$

which is non-zero for $l = k \pm 1$. [See Eq. (B.9).] We also have

$$\langle t|\partial/\partial X|s\rangle_2 = [(1-\lambda^2)/\beta]^{1/4}\langle t|\partial/\partial z|s\rangle_0, \tag{63}$$

which is non-zero for $t = s \pm 1$. Hence Eq. (46) becomes

$$\langle\, tl \,|\, H \,|\, sk \,\rangle = -\lambda\sqrt{\beta B/A}\;\langle l|\partial/\partial\xi|k\rangle_0\;\langle t|\partial/\partial z|s\rangle_0. \tag{64a}$$

Next calculate the non-adiabatic corrections to the energy using Rayleigh-Schrödinger perturbation theory (RSPT). There is no first-order correction, since all diagonal corrections are included in $E_{sk}^{0}$, Eq. (61a). The second-order correction is given by the usual formula,[1]

$$
\begin{aligned}
E_{sk}^{(2)} &= -\sum_{l = k \pm 1} \sum_{t = s \pm 1} |\langle \, tl \mid H \mid sk \, \rangle|^2 (E_{tl}^{0} - E_{sk}^{0})^{-1} \\
&= -\beta\lambda^2 (B/A) \sum_{l,\,t} \frac{|\langle l | \partial / \partial\xi | k \rangle_0|^2 |\langle t | \partial / \partial z | s \rangle_0|^2}{(l-k)A + (t-s)B}
\end{aligned}
\quad . \tag{65}
$$

The final result, after working out the algebra, is

$$
E_{sk}^{(2)} = \tfrac{1}{2}\beta\lambda^2 (B/A)[(k + \tfrac{1}{2})B - (s + \tfrac{1}{2})A]/(A^2 - B^2) \tag{66}
$$

The second-order non-adiabatic results, $E_{sk}^{0} + E_{sk}^{(2)}$, are shown by dashed blue lines in Figs. 1 and 2. The corrections markedly improve agreement with the exact results, although some deviations are still seen for large values of $\beta$, particularly for the individual oscillator terms.

More accurate results can be obtained by calculating the non-adiabatic corrections to all orders, using the numerical perturbation theory methods discussed in Ref. [35]. This part of the calculation is not trivial, even for such a simple problem, because of degeneracy effects.[2] The "iterative" method [35] was used here because it was much more robust than the RSPT method. Further improvements could be obtained using the "synthetic Hamiltonian" method [35] or by matrix diagonalization, but those refinements were not needed for the purposes of this work.

The results obtained using the iterative method are shown by the green curves in Figs. 1 and 2. Agreement with the exact results is virtually perfect.

Solution of this simple problem is, of course, not an end in itself. I have included it only to demonstrate the methodology. Discussion of the "real" electron-nuclear problem resumes in the next section.

---

1. Note that RSPT can be applied even though the basis functions are not eigenfunctions of a Hamiltonian operator and the perturbation can only be expressed in terms of the matrix elements. Also note that the diagonal terms do not have to be included in the perturbation expansion and can be used in the energy denominators that arise in the expansion. See Ref. [35] for discussion of these issues.
2. For example, note the energy of the $k = 1$, $s = 0$ state is close to that of the $k = 0$, $s = 3$ state, for $\lambda = 0.9$, $\beta = 1$. These two states are not directly coupled through matrix elements, but they are coupled through intermediate states, so that the degeneracy effects appear in high order.

# 6. CALCULATION OF THE ELECTRONIC FUNCTIONS

The xiabatic electronic wavefunctions and energy surfaces are defined by Eqs. (36) and (38). I have not yet found an exact solution to these equations or even proved that a solution exists. This section outlines an approximate solution that should be useful, at least as a starting point.

## 6.1 Preliminary Observations

The xiabatic matrix $\langle l\,|H_e^0|\,k\rangle_e + \nu_{lk}$, cannot be diagonalized by standard methods. Suppose one attempts to expand the exact electronic functions in an approximate basis set $\Phi_k^0$, using a unitary transformation

$$\Phi_k(\vec{x};\vec{X}) = \Sigma_m U_{mk}(\vec{X})\Phi_m^0(\vec{x};\vec{X}), \text{ where } U_{mk}^*(\vec{X}) = U_{km}^{-1}(\vec{X}). \quad (67)$$

Since the expansion coefficients $U_{mk}$ are functions of the nuclear coordinates, the $\nu_{lk}$ terms involve derivatives $\nabla_\alpha U_{mk}$. Hence one cannot satisfy Eq. (38) simply by finding the $U_{mk}$ that diagonalize $\langle l\,|H_e^0|\,k\rangle_e^0 + \nu_{lk}^0$.

The strategy employed here will be to replace as much of the $\nu_{lk}$ term as possible with quantities that do not depend on the $\nabla_\alpha$ derivatives. This procedure will give an effective Hamiltonian and eigenfunctions.

Also note that all terms in the xiabatic matrix are independent of the origin and orientation of the coordinate system. The $\nu_{lk}$ terms depend only on the *topology* of the wavefunctions and potential surface at each configuration. The method outlined below does employ a coordinate transformation, but not a shift in the origin or rotation of the axes.

## 6.2 Definitions

The total numbers of electrons and nuclei are given by

$$N_e = \Sigma_i, \quad N = \Sigma_\alpha. \quad (68)$$

Also define

$$\overline{M} = N^{-1}\Sigma_\alpha M_\alpha, \quad \mu_\alpha = M_\alpha/\overline{M}, \quad \varepsilon = m_e/\overline{M}, \quad \kappa = \hbar^2/2m_e. \quad (69)$$

Here $\overline{M}$ is the average nuclear mass, $\mu_\alpha$ are reduced masses, and $\varepsilon$ is the electron-nuclear mass ratio. Note that $\Sigma_\alpha\mu_\alpha = N$.

The electronic kinetic energy operator, from Eq. (3), is

$$K_e(\vec{x}) = -\kappa\sum_i \nabla_i^2 . \tag{70}$$

The xiabatic correction terms, from Eq. (36), are

$$\nu_{lk} = \varepsilon\kappa\sum_\alpha \mu_\alpha^{-1}\int(\nabla_\alpha\Phi_l^*)\cdot(\nabla_\alpha\Phi_k)d\vec{x} \tag{71}$$

## 6.3 Coordinate Transformation

Consider the following coordinate transformation:

$$\boldsymbol{q}_i = \boldsymbol{x}_i + aN_e^{-1}\Sigma_j\boldsymbol{x}_j, \quad \boldsymbol{Q}_\alpha = (1+a)\boldsymbol{X}_\alpha, \tag{72}$$

where $a$ is a constant to be determined. The reverse transformation is

$$\boldsymbol{x}_i = \boldsymbol{q}_i + c\boldsymbol{q}_e, \quad \boldsymbol{q}_e = N_e^{-1}\Sigma_j\boldsymbol{q}_j, \quad \boldsymbol{X}_\alpha = (1+c)\boldsymbol{Q}_\alpha, \text{ where} \tag{73a}$$

$$c = -a/(1+a), \quad a = -c/(1+c). \tag{73b}$$

The gradient operators transform as follows.

$$\nabla_i = \nabla_{qi} + a\boldsymbol{G}_e, \quad \boldsymbol{G}_e = N_e^{-1}\Sigma_j\nabla_{qj}, \quad \nabla_\alpha = (1+a)\nabla_{q\alpha}, \tag{74}$$

where $\nabla_{qi}$ and $\nabla_{q\alpha}$ denote derivatives with respect to the new coordinates $\boldsymbol{q}_i$ and $\boldsymbol{Q}_\alpha$, respectively.

Integrals over the $\boldsymbol{x}_i$ can be expressed as integrals over the $\boldsymbol{q}_i$, and vice versa. The Jacobian for the transformation [43] is derived in Appendix C. If $\Theta$ is some function of the $\boldsymbol{x}_i$, Eq. (C.7) gives

$$\int\Theta(\vec{x})\, d\vec{x} = (1+c)^3\int\Theta[\vec{x}(\vec{q})]\, d\vec{q}. \tag{75}$$

In what follows, the expression $\langle l|\Omega|k\rangle_e$ will denote matrix elements of an operator in which the integration is taken over the $\boldsymbol{x}_i$, not over the $\boldsymbol{q}_i$.

## 6.4 Transformation of the Hamiltonian

The electronic kinetic energy operator, in terms of the new coordinates, is

$$K_e(\vec{x}) = K_e(\vec{q}) - \kappa(2a + a^2)N_e G_e^2. \tag{76}$$

The potential energy terms $U_{ee}$ and $U_{nn}$, Eq. (4), transform as follows.

$$U_{ee}(\vec{x}) = \sum_{i>j} e^2/|\boldsymbol{q}_i - \boldsymbol{q}_j| = U_{ee}(\vec{q}), \tag{77a}$$

$$U_{nn}(\vec{X}) = (1+a)\sum_{\alpha>\beta} Z_\alpha Z_\beta\, e^2/|\boldsymbol{Q}_\alpha - \boldsymbol{Q}_\beta| = (1+a)U_{nn}(\vec{Q}). \tag{77b}$$

The electron-nuclear potential, in terms of the new coordinates, is given by Eq. (5a), with

$$r_{i\alpha} = |\boldsymbol{x}_i - \boldsymbol{X}_\alpha| = |\boldsymbol{q}_i - \boldsymbol{Q}_\alpha + (1+c)(\boldsymbol{q}_e - \boldsymbol{Q}_\alpha)|. \tag{78}$$

As shown below, the constant $c$ is very small, proportional to the mass ratio $\varepsilon$. Therefore, it will be a good approximation to expand $U_{en}$ in powers of $c$, retaining only the first two terms,

$$U_{en}(\vec{x}, \vec{X}) = U_{en}(\vec{q}, \vec{Q}) + c(\partial U_{en}/\partial c)_0 + O(c^2). \tag{79}$$

The first-order term can be written in the form

$$(\partial U_{en}/\partial c)_0 = \tfrac{1}{2}\sum_i \sum_j \chi(\boldsymbol{q}_i, \boldsymbol{q}_j; \vec{Q}) = V_{ee}(\vec{q}, \vec{Q}), \tag{80a}$$

$$\chi(\boldsymbol{q}_i, \boldsymbol{q}_j; \vec{Q}) = N_e^{-1}\sum_j Z_\alpha e^2(\boldsymbol{q}_i - \boldsymbol{Q}_\alpha)\cdot(\boldsymbol{q}_j - \boldsymbol{Q}_\alpha)(\eta_{i\alpha}^{-3} + \eta_{j\alpha}^{-3}), \tag{80b}$$

where $\eta_{i\alpha} = max(|\boldsymbol{q}_i - \boldsymbol{Q}_\alpha|, r_{0\alpha})$. Hence $V_{ee}$ has the form of an electron-electron interaction term, in which the electrons are coupled through their interactions with the nuclei.

The xiabatic correction terms are

$$\nu_{lk} = \varepsilon\kappa(1+a)\sum_\alpha \mu_\alpha^{-1}\int(\nabla_{q\alpha}\Phi_l^*)\cdot(\nabla_{q\alpha}\Phi_k)d\vec{x}. \tag{81}$$

## 6.5 Electronic Wavefunctions and Energies

Now define a set of electronic wavefunctions and energy surfaces $\Phi_k^0$ by

$$H_e(\vec{q};\vec{Q})\,\Phi_k^0(\vec{q};\vec{Q}) = W_k^0(\vec{Q})\,\Phi_k^0(\vec{q};\vec{Q}), \tag{82}$$

where $H_e$ is the effective Hamiltonian

$$\begin{aligned} H_e(\vec{q};\vec{Q}) &= K_e(\vec{q}) + U_{en}(\vec{q},\vec{Q}) + U_{ee}(\vec{q}) \\ &\quad + cV_{ee}(\vec{q},\vec{Q}) + (1+a)U_{nn}(\vec{Q}) \end{aligned} . \tag{83}$$

Hence $H_e$ is essentially the clamped-nuclei Hamiltonian, in which the $\boldsymbol{x}_i$ have been replaced by the $\boldsymbol{q}_i$, plus the new electron-electron interaction term $V_{ee}$.

The energy surface $W_k^0$ obtained by this equation can be expressed as functions of either the $\boldsymbol{X}_\alpha$ or the $\boldsymbol{Q}_\alpha$ and will have the same form in both cases, apart from the small scaling factor $1 + a$.

Transformation of the wavefunctions from the $\boldsymbol{q}_i$ back to the original coordinates $\boldsymbol{x}_i$ will lead to more significant changes in the nature of the functions. One must also be careful about normalization of the wavefunctions. From Eq. (75), the orthogonality relation will be

$$\int \Phi_l^{0*} \Phi_k^0 \, d\vec{x} = \delta_{lk} = (1+c)^3 \int \Phi_l^{0*} \Phi_k^0 \, d\vec{q} . \tag{84}$$

If, in the calculation process, the wavefunctions are chosen to be orthonormal in terms of the $\boldsymbol{q}_i$, they must be multiplied by the factor $(1+c)^{-3/2}$ to satisfy this condition.

The reason for reformulating the problem in terms of the new coordinates is that it allows one to replace much of the xiabatic corrections $\nu_{lk}$ with different expressions, as shown below.

## 6.6 Cancellation of Terms

The matrix elements of the original CN Hamiltonian, in terms of the new electronic wavefunctions, are

$$\langle l | H_e^0(\vec{x};\vec{X}) | k \rangle_e^0 = \delta_{lk} W_k^0 + \kappa(2a + a^2) N_e \int \boldsymbol{G}_e \Phi_l^{0*} \cdot \boldsymbol{G}_e \Phi_k^0 d\vec{x} . \tag{85}$$

The next step is to choose the constant $a$ to cancel some of the xiabatic correction terms $\nu_{lk}$. Define the following operators:

$$\boldsymbol{G}_n = N^{-1} \Sigma_\alpha \nabla_{q\alpha}, \quad \boldsymbol{F}_\alpha = \nabla_{q\alpha} - \mu_\alpha \boldsymbol{G}_n, \quad F = \boldsymbol{G}_n + \boldsymbol{G}_e . \tag{86}$$

After some manipulation, one obtains

$$\nu_{lk}^0 = \Delta\nu_{lk}^0 + \xi_{lk}^0 + \varepsilon\kappa(1+a)N\int \boldsymbol{G}_e\Phi_l^{0*} \cdot \boldsymbol{G}_e\Phi_k^0 d\vec{\boldsymbol{x}}, \text{ where} \tag{87a}$$

$$\Delta\nu_{lk}^0 = \varepsilon\kappa(1+a)\sum_\alpha \mu_\alpha^{-1}\int \boldsymbol{F}_\alpha\Phi_l^{0*} \cdot \boldsymbol{F}_\alpha\Phi_k^0 d\vec{\boldsymbol{x}}, \text{ and} \tag{87b}$$

$$\xi_{lk}^0 = (1+a)\int[\boldsymbol{F}\Phi_l^{0*}\cdot \boldsymbol{F}\Phi_k^0 + \boldsymbol{F}\Phi_l^{0*}\cdot \boldsymbol{G}_e\Phi_k^0 + \boldsymbol{G}_e\Phi_l^{0*}\cdot \boldsymbol{F}\Phi_k^0]d\vec{\boldsymbol{x}}\ . \tag{87c}$$

In Appendix A, it is shown that $\xi_{lk}^0 = 0$ if the wavefunctions are real.[1]

Now add Eqs. (85) and (87a) to obtain an expression for $\langle l|H_e^0|k\rangle_e^0 + \nu_{lk}^0$. The integrals involving $\boldsymbol{G}_e\Phi_l^{0*}\cdot\boldsymbol{G}_e\Phi_k^0$ can be made to cancel by requiring

$$(2a + a^2)N_e + \varepsilon(1+a)N = 0, \quad a = \sqrt{1 + (\varepsilon N/N_e)^2} - 1 - \varepsilon N/2N_e\ . \tag{88}$$

Since $\varepsilon$ is very small, it will be a good approximation to expand $a$ in powers of $\varepsilon$ and keep only the linear term. Hence Eq. (73b) becomes

$$a \cong -\varepsilon N/2N_e, \qquad c \cong \varepsilon N/2N_e. \tag{89}$$

Using the above results, one finds the following results for the diagonal and off-diagonal elements of the matrix,

$$\langle k|H_e^0|k\rangle_e^0 + \nu_{kk}^0 = W_k^0(\vec{\boldsymbol{Q}}) + \Delta\nu_{kk}^0 = W_k(\vec{\boldsymbol{X}})\ , \tag{90a}$$

$$\langle l|H_e^0|k\rangle_e^0 + \nu_{lk}^0 = \Delta\nu_{lk}^0 \text{ for } l \neq k. \tag{90b}$$

Here $W_k(\vec{\boldsymbol{X}})$ will be the potential energy surface for a calculating the nuclear wavefunctions, Eq. (44).

## 6.7 Discussion

The procedure outlined above does not completely diagonalize the xiabatic matrix $\langle l|H_e^0|k\rangle_e + \nu_{lk}$, but it should substantially reduce the magnitude of the off-diagonal elements. Examination of the formula for the $\Delta\nu_{lk}^0$, Eq. (87b), suggests that these terms will be small, certainly smaller than the $\nu_{lk}^0$. Since $\sum_\alpha \boldsymbol{F}_\alpha = 0$, the op-

1. The proof given in Appendix A is strictly applicable to nuclear configurations where the potential surfaces are non-degenerate. I assume here that the result holds for degenerate points if it holds for all configurations around it. I admit that a more rigorous proof would be desirable.

erator $\boldsymbol{F}_\alpha$ is a kind of "fluctuation" in the derivative $\nabla_{q\alpha}$ about its average over all nuclear coordinates. These fluctuations cannot be expected to cancel exactly, but it may be possible to ignore the off-diagonal terms. If not, they can be added to the non-adiabatic terms in Eq. (46).

It should be possible to apply this procedure to existing calculations on specific systems for which CN wavefunctions and potential surfaces already exist. For potential surfaces that are well-isolated, these corrections should be quite small. In such cases the $V_{ee}$ term could be treated by first-order perturbation theory, which affects only the energy and not the wavefunctions. In the neighborhood of degenerate points, however, these corrections might require a linear combination of CN wavefunctions to diagonalize the $V_{ee}$ matrix elements, as is done in transforming to the diabatic basis set.

This point raises the question: What effect does use of the xiabatic wavefunctions have on the geometric phase effect and related singularities? It is well-known that eigenfunctions of the CN Hamiltonian have an arbitrary phase factor [22]: If $\Phi_k$ is an eigenfunction of $H_e^0$, then $\Phi_k \exp(if_k)$, where $f_k$ is a function only of the nuclear coordinates, is also an eigenfunction. This non-uniqueness gives rise to the geometric phase effect, where transport of the wavefunction around a closed loop, containing a singularity, leads to a change of phase.

Note that his non-uniqueness does not occur when the $\nu_{lk}$ terms are included. Suppose one has wavefunctions $\Phi_k^0$ that are *exact* solutions to Eq. (38). Now define new functions that differ only by a phase factor, $\Phi_k = \Phi_k^0 \exp(if_k)$. The xiabatic matrix, in terms of the new functions, is

$$\langle l \,|H_e^0|\, k\rangle_e + \nu_{lk} = \delta_{lk}(W_k^0 + \varepsilon\kappa\Sigma_\alpha|\nabla_\alpha f_k|^2) + i\omega_{lk}(f_k + f_l)e^{i(f_k - f_l)} . \qquad (91)$$

Hence the phase shifts $f_k$ modify the diagonal elements and also generate new off-diagonal elements.[1]

Further exploration of this and other questions will be appropriate when the method is applied to specific problems.

Finally, note that the electron-nuclear coupling terms, Eq. (29), are given by

$$\omega_{lk} = -\varepsilon\kappa(1 + a)\Sigma_\alpha\langle l \,|\nabla_{q\alpha}|\, k\rangle_e \cdot \nabla_\alpha . \qquad (92)$$

1. This statement excludes the possibility of both deliberate and fortuitous cancellations of the phase-shift terms. Also note that the new terms vanish if the $f_k$ are constants, not functions of the nuclear coordinates.

Using the results obtained above, this equation can also be written

$$\omega_{lk} = -\varepsilon\kappa(1+a)[\Sigma_\alpha\langle l\,|\boldsymbol{F}_\alpha|\,k\rangle_e^0 \cdot \nabla_\alpha - \langle l\,|\boldsymbol{G}_e|\,k\rangle_e^0 \cdot \Sigma_\alpha \mu_\alpha \nabla_\alpha]. \qquad (93)$$

Hence much of the above term can be calculated without having to compute the derivatives of the wavefunctions with respect to the nuclear coordinates. The above expression may also be useful in calculating the nuclear wavefunctions and electron-nuclear coupling matrix elements.

# 7. SUMMARY AND CONCLUSIONS

This report presents a substantially new approach to the treatment of electron-nuclear coupling in quantum theories of molecules and condensed matter. It employs a perturbation method in which the traditional "Born-Oppenheimer approximation"[1] is the leading term. The method can be used both to identify cases where the BOA is sufficiently accurate and also to calculate corrections when it is not. The corrections can be calculated to as high a degree of accuracy as desired, at least in principle.

A central feature of the method is to separate the corrections into two types—adiabatic and non-adiabatic.

- The adiabatic terms use a single electronic potential surface for the nuclear motion but go beyond the "clamped-nuclei" solutions to produce an extended, or "xiabatic," basis set of wavefunctions and energies. An approximate solution to this problem is outlined in Sec. 6.
- The non-adiabatic terms allow the nuclei to move on more than one potential surface. In order to compute these corrections, one must compute certain electron-nuclear coupling matrix elements and then apply either perturbation theory or matrix diagonalization methods.

The perturbation method has been demonstrated, in Sec. 5, on the problem of two coupled harmonic oscillators. This simple example is quite different from that of a molecule or condensed matter, but it does illustrate the mechanics of the method and shows how a nominally exact result can be obtained by computing the perturbation corrections to all orders.

Further developments of the theory will be facilitated by its application to specific problems of interest. It should be possible to use the theory to analyze numerical results that are already available. As noted in Sec. 6, the xiabatic wavefunctions and potential surfaces can be calculated from existing clamped-nuclei solutions using coordinate transformations and linear combinations. Exploratory studies on the simplest ssytems would be most fruitful at the outset. They may identify areas where improvements are needed and suggest ways to make changes.

One of the most important issues to be investigated is the effect of the xiabatic corrections on the problem of the geometric phase shift and other singularities. As noted in Sec. 6.7, the xiabatic wavefunctions cannot have an arbitrary phase factor as do the clamped-nuclei wavefunctions. The extent to which the xiabatic formu-

1. Once again, this term is used in a number of different ways throughout the literature. I am using the most common definition, given at the beginning of Sec. 1.1.

lation corrects these problems needs to be investigated for specific cases, especially when the approximate solution given in Sec. 6 is used.

The method developed here was intended primarily for the study of electrons and nuclei interacting by Coulomb potentials. However, many of the results will be equally applicable to calculations using approximate potentials and wavefunctions, such as density-functional methods, which are more common in studies of condensed matter. The method should also be useful for other kinds of coupling problems, such as coupling between the translational, vibrational, and rotational degrees of freedom in solids and liquids.

Note added, June 24, 2013: I do not claim that the ideas and methods presented in this report represent a complete solution to the problem of calculating corrections to the Born-Oppenheimer approximation. Further developments and refinements will be needed before they can be applied to real problems. However, I *do* believe that they represent a significant improvement on what has been done in the past and offer opportunities that can be exploited by those willing to consider them with an open mind.

# Appendix A

## Relations Involving Coulomb Potentials

This appendix gives additional details about relations presented in the body of the report.

### Finite-Nucleus Correction

The standard Coulomb expression for the electron-nuclear potential $U_{en}$, which treats both particles as point charges, gives rise to certain numerical problems. In particular, the electronic wavefunctions have cusps at the nuclear positions, $\boldsymbol{x}_i = \boldsymbol{X}_\alpha$, where second derivatives with respect to the electronic and nuclear positions are undefined. These problems can be eliminated by accounting for the finite sizes of the nuclei.

This appendix presents a correction to the potential derived by assuming the nuclei to be spherically symmetric and have a uniform charge density, while the electrons are still treated as point charges.[1] A typical nuclear radius, accurate to about 10%, is $r_0 = 2.3 \times 10^{-5} A^{1/3}\,\mathrm{a}_0$, where $A$ is the mass number and $\mathrm{a}_0$ is the Bohr radius ($0.5292 \times 10^{-5}$ cm) [32].

Let $u_{en}(r)$ be the interaction potential between an electron at a radius $r$ from the center of a nucleus with charge $Z$ and radius $r_0$. This function can be calculated using standard methods of electrostatics [33][34]. The correction only affects the potential when the electron is inside the nuclear sphere, and the result is

$$u_{en}(r) = \begin{bmatrix} -Ze^2/r \text{ for } r \geq r_0, \\ -Ze^2[3-(r/r_0)^2]/2r_0 \text{ for } r \leq r_0 \end{bmatrix} \tag{A.1}$$

When this correction is included, the electron-nuclear potential term is

$$U_{en} = -\sum_i \sum_\alpha Z_\alpha e^2 f(r_{i\alpha}/r_{0\alpha})/r_{i\alpha}, \text{ where } r_{i\alpha} = |\boldsymbol{x}_i - \boldsymbol{X}_\alpha|, \tag{A.2}$$

1. In principle, one could also make a correction for the electron size as well. However, there are several ways to define the electron radius, and it is not clear which would be appropriate for present purposes. Taking the ratio of the electron and proton radii equal to the cube root of their mass ratio, the electron radius would be 8% of the proton radius. That would give only a small refinement to the formula derived here.

$$f(x) = 1 \text{ for } x \geq 1, \text{ and } f(x) = (3x - x^3)/2 \text{ for } x \leq 1. \tag{A.3}$$

## Derivatives

**Electron-Electron Potential**. The first derivatives of $U_{ee}$ are

$$\nabla_i U_{ee} = -\sum_{j \neq i} e^2(\boldsymbol{x}_i - \boldsymbol{x}_j)/r_{ij}^3. \tag{A.4}$$

The matrix elements of $\nabla_i U_{ee}$, are found to vanish because of symmetry with respect to interchange of two electrons:

$$\langle l|(\boldsymbol{x}_i - \boldsymbol{x}_j)/r_{ij}^3\ |k\rangle_e = \langle l|(\boldsymbol{x}_j - \boldsymbol{x}_i)/r_{ij}^3\ |k\rangle_e = 0, \tag{A.5}$$

$$\langle l|\nabla_i U_{ee}|k\rangle_e = -\langle l|\nabla_i U_{ee}|k\rangle_e = 0. \tag{A.6}$$

The second derivatives of $U_{ee}$ consist of Dirac delta-functions.

$$\nabla_i^2 U_{ee} = \sum_{j \neq i} e^2 \delta(\boldsymbol{x}_i - \boldsymbol{x}_j). \tag{A.7}$$

**Electron-Nuclear Potential**. The first derivatives of $U_{en}$ are

$$\nabla_\alpha U_{en} = -Z_\alpha e^2 \sum_i (\boldsymbol{x}_i - \boldsymbol{X}_\alpha)/\eta_{i\alpha}^3, \tag{A.8}$$

$$\nabla_i U_{en} = \sum_\alpha Z_\alpha e^2 (\boldsymbol{x}_i - \boldsymbol{X}_\alpha)/\eta_{i\alpha}^3, \tag{A.9}$$

where $\eta_{i\alpha} = max(r_{i\alpha}, r_{0\alpha})$.

The following result is useful.

$$\sum_\alpha \nabla_\alpha U_{en} = -\sum_i \nabla_i U_{en}. \tag{A.10}$$

The second derivatives of $U_{en}$ are

$$\nabla_\alpha^2 U_{en} = (3 Z_\alpha e^2 / r_{0\alpha}^3) \Sigma_{\alpha i}, \tag{A.11}$$

$$\nabla_i^2 U_{en} = \Sigma_{\alpha i} Z_\alpha (3e^2/r_{0\alpha}^3), \tag{A.12}$$

where the sums $\Sigma_{\alpha i}$ are taken over all electron-nuclear pairs for which $r_{i\alpha} < r_{0\alpha}$.

Note that, in the limit $r_0 \to 0$, $\nabla^2_\alpha U_{en}$ and $\nabla^2_i U_{en}$ become delta-functions, as in Eq. (A.7), and it is this property that gives rise to the cusps in the wavefunctions. Equation (A.2) spreads the delta-functions over small but finite spheres and reduces the probability of an electron being inside. That also eliminates the cusps.

It is true that $\nabla^2_\alpha U_{en}$ and $\nabla^2_i U_{en}$ are discontinuous at the boundaries of the nuclei, $r_{i\alpha} = r_{0\alpha}$, where the higher derivatives do not exist. Those derivatives do not appear to be needed for the methods discussed in this report. Even if they were needed, they could be made to exist by making small changes to the function $f$, defined in Eq. (A.3).

**Nuclear-Nuclear Potential**. The derivatives of $U_{nn}$ are analogous to those for $U_{ee}$. They are not needed in the present work.

## Matrix Elements

Now consider the calculation of matrix elements using the electronic functions defined in Sec. 6.5. The $\Phi^0_k$ are eigenfunctions of the effective Hamiltonian $\boldsymbol{H}_e$, Eq. (83), which is expressed as a function of the coordinates $\boldsymbol{q}_i$ and $\boldsymbol{Q}_\alpha$, defined in Sec. 6.3. The relations derived above also apply to these coordinates.

The commutator theorem, Eq. (23), gives the following equations, both of which are forms of the Hellmann-Feynman theorem:

$$(W^0_k - W^0_l)\langle l|\nabla_{q\alpha}|k\rangle^0_e = \langle l|\nabla_{q\alpha}(U_{en} + cV_{ee})|k\rangle^0_e + \delta_{kl}\nabla_{q\alpha}(U_{nn} - W^0_k), \quad (A.13)$$

$$(W^0_k - W^0_l)\langle l|\nabla_{qi}|k\rangle^0_e = \langle l|\nabla_{qi}(U_{en} + cV_{ee} + U_{ee})|k\rangle^0_e, \quad (A.14)$$

where the term $\langle l|\nabla_{qi}U_{ee}|k\rangle^0_e$ drops out from Eq. (A.6).

**The $\boldsymbol{F}$ Operator**. Consider the matrix elements of the operator $\boldsymbol{F} = \boldsymbol{G}_n + \boldsymbol{G}_e$. Using Eqs. (74), (86), (A.13), and (A.14), one finds

$$(W^0_k - W^0_l)\langle l|\boldsymbol{F}|k\rangle^0_e = \langle l|\boldsymbol{F}(U_{en} + cV_{ee})|k\rangle^0_e. \quad (15)$$

The contribution from $U_{en}$ vanishes from Eq. (A.10). Inspection of Eqs. (80a) and (80b) shows that a similar relation holds for $V_{ee}$,

$$\Sigma_\alpha \nabla_{q\alpha} V_{ee} = -\Sigma_i \nabla_{qi} V_{ee}. \quad (A.16)$$

Hence

$$\langle l|\boldsymbol{F}|k\rangle_e^0 = 0 \text{ for } l \neq k\,. \qquad \text{(A.17)}$$

Strictly speaking, the above result only applies either to non-degenerate configurations, $W_k^0 \neq W_l^0$, or to cases where the matrix element can be set to zero for other reasons, e.g., differences in spin between the two states. For the present, however, I will assume that, if Eq. (A.16) holds in the neighborhood of a degenerate point, it will also hold at the degenerate point. This issue may require further study.

The diagonal elements of $\boldsymbol{F}$ can also be made to vanish. It was previously shown, in Eq. (30), that $\langle l|\nabla_\alpha|k\rangle_e$ is anti-Hermitan, and the same argument applies to $\langle l|\nabla_{qi}|k\rangle_e^0$ and $\langle l|\nabla_{q\alpha}|k\rangle_e$. If the wavefunctions are real, the diagonal matrix elements must vanish. Assuming this condition to be hold,

$$\langle k|\boldsymbol{F}|k\rangle_e^0 = 0 \text{ if } \Phi_k^0 \text{ is real.} \qquad \text{(A.18)}$$

**The $\xi_{lk}^0$ Matrix elements**. Now apply this result to the term $\xi_{lk}^0$, Eq. (87c). Making use of the sum rule, Eq. (19), one finds

$$\int \boldsymbol{F}\Phi_l^{0*} \cdot \boldsymbol{F}\Phi_k^0 d\hat{\boldsymbol{x}} = \Sigma_m \langle l|\boldsymbol{F}|m\rangle_e^0 \cdot \langle m|\boldsymbol{F}|k\rangle_e^0 = 0\,. \qquad \text{(A.19)}$$

The other two terms in Eq. (87c) are found to vanish by the same argument—apply the sum rule and note that all terms in the sum involve matrix elements $\langle l|\boldsymbol{F}|m\rangle_e^0$, which vanish. Hence

$$\xi_{lk}^0 = 0\,, \text{Q.E.D.} \qquad \text{(A.20)}$$

# Appendix B

## Harmonic Oscillator Relations

This appendix presents formulas that are used in solving the coupled-oscillator example, Sec. 5. See Refs. [35] and [42] for additional details.

The one-dimensional harmonic oscillator Hamiltonian, in dimensionless units, is

$$h_0(\xi) = -\tfrac{1}{2}\,\partial^2/\partial\xi^2 + \tfrac{1}{2}\xi^2. \tag{B.1}$$

The eigenvalues and eigenfunctions of $h_0$ are

$$h_0(\xi)\psi_k(\xi) = (k+\tfrac{1}{2})\psi_k(\xi), \tag{B.2}$$

$$\psi_k(\xi) = N_k P_k(\xi) e^{-\xi^2/2}, \quad N_k = [\pi^{1/4}(2^k k!)^{1/2}]^{-1}, \tag{B.3}$$

where $P_k(\xi)$ is a Hermite polynomial. The normalization constant $N_k$ is chosen so that $\psi_k(\xi)$ is normalized on the interval $-\infty \le \xi \le \infty$. (Note that different factors apply to the wavefunctions of Sec. 5, where the independent variables differ from $\xi$ by constant factors.)

The Hermite polynomials satisfy the well-known relationships

$$P_{k+1} - 2\xi P_k + 2kP_{k-1} = 0 \text{ (with } P_0 = 1\text{), and} \tag{B.4}$$

$$dP_k/d\xi = 2kP_{k-1} = 2\xi P_k - P_{k+1}. \tag{B.5}$$

Hence the wavefunctions $\psi_k$ satisfy the relationships

$$\xi\psi_k = \sqrt{(k+1)/2}\;\psi_{k+1} + \sqrt{k/2}\;\psi_{k-1}\text{, and} \tag{B.6}$$

$$d\psi_k/d\xi = \xi\psi_k - \sqrt{2(k+1)}\psi_{k+1}. \tag{B.7}$$

Then the only non-zero matrix elements $\langle l|\xi|k\rangle$ are

$$\langle k+1|\xi|k\rangle = \langle k|\xi|k+1\rangle = \sqrt{(k+1)/2}. \tag{B.8}$$

The only non-zero matrix elements $\langle l|d/d\xi|k\rangle$ are

$$\langle k+1|d/d\xi|k\rangle = -\langle k|d/d\xi|k+1\rangle = -\sqrt{(k+1)/2}\,. \tag{B.9}$$

The matrix elements $\langle l|\xi^2|k\rangle$ can be determined from the sum rule, Eq. (19). The only non-zero values are

$$\langle k|\xi^2|k\rangle = k+\tfrac{1}{2}, \tag{B.10}$$

$$\langle k|\xi^2|k+2\rangle = \langle k+2|\xi^2|k\rangle = \tfrac{1}{2}\sqrt{(k+1)(k+2)}\,. \tag{B.11}$$

Also note that

$$\tfrac{1}{2}\int_{-\infty}^{\infty}\frac{d\psi_l}{d\xi}\frac{d\psi_k}{d\xi}d\xi = -\tfrac{1}{2}\int_{-\infty}^{\infty}\psi_l\, d^2\psi_k/d\xi^2 d\xi = \langle l|h_0|k\rangle - \tfrac{1}{2}\langle l|\xi^2|k\rangle\,. \tag{B.12}$$

# Appendix C

## Jacobian of the Transformation

Consider the coordinate transformation given by Eqs. (73a). The volume element associated with this transformation is given by [43]

$$d\tilde{x}_1 d\tilde{x}_2 \ldots = J_x J_y J_z d\tilde{q}_1 d\tilde{q}_2 \ldots, \tag{C.1}$$

where $J_x$, $J_y$, and $J_z$ are the Jacobian determinants for the transformation. (Note that $d\tilde{x}_i$ and $d\tilde{q}_i$ each correspond to three orthogonal axes, so that one obtains a determinant for each direction.)

Consider the Jacobian $J_y$, which is formed from the derivatives

$$dy_i / dq_{yj} = \delta_{ij} + c / N_e = \delta_{ij} + \tau. \tag{C.2}$$

Hence

$$J_y = \begin{vmatrix} 1+\tau & \tau & \ldots & \tau \\ \tau & 1+\tau & \ldots & \tau \\ \ldots & \ldots & \ldots & \ldots \\ \tau & \tau & \ldots & 1+\tau \end{vmatrix}, \tag{C.3}$$

which consists of $N_e$ rows and columns.

$J_y$ can be calculated as follows: The value of a determinant is unchanged if any row is added to any other row [43]. In Eq. (C.3), sum all rows, starting with the second, and add the result to the first row. The result is

$$J_y = \begin{vmatrix} 1+c & 1+c & \ldots & 1+c \\ \tau & 1+\tau & \ldots & \tau \\ \ldots & \ldots & \ldots & \ldots \\ \tau & \tau & \ldots & 1+\tau \end{vmatrix} = (1+c) \begin{vmatrix} 1 & 1 & \ldots & 1 \\ \tau & 1+\tau & \ldots & \tau \\ \ldots & \ldots & \ldots & \ldots \\ \tau & \tau & \ldots & 1+\tau \end{vmatrix}, \tag{C.4}$$

where the factor $1 + c$ can be brought outside the determinant because it is identical for all elements on the first row.

Now multiply the first row by $\tau$ and subtract it from each of the other rows. This operation yields

$$J_y = (1+c)\begin{vmatrix} 1 & 1 & \dots & 1 \\ 0 & 1 & \dots & 0 \\ \dots & \dots & \dots & \dots \\ 0 & 0 & \dots & 1 \end{vmatrix} = 1+c, \tag{C.5}$$

where the final step is obtained by evaluating the determinant by its first column.

Since all three Jacobians are identical,

$$J_x = J_y = J_z = 1+c. \tag{C.6}$$

Hence the transformation of the volume element, Eq. (C.1), is

$$\boxed{d\vec{x}_1 d\vec{x}_2 \dots = (1+c)^3 d\vec{q}_1 d\vec{q}_2 \dots}. \tag{C.7}$$

(Note that this result is independent of the number of electrons $N_e$.)